# Bodyless Presence:

## Reconsidering the Minimal Self in Immersive Video

**TOIDA Koichi**
MESON, Inc.
koichi.toida@meson.tokyo

**ABSTRACT** Immersive video, namely 180-degree and 360-degree video designed to be viewed through head-mounted displays, constitutes an important boundary case between interactive VR and conventional two-dimensional video viewing for reconsidering self-experience in XR. In immersive video, the user can select the direction of the viewpoint through head rotation, while being unable to actively change the recorded environment through walking, approaching, grasping, or manipulating. In many cases, no explicit body or avatar corresponding to the user is provided. This paper reinterprets presence in immersive video not as bodily extension or body ownership of an avatar, but as a form of self-experience in which self-location becomes relatively dominant under conditions of reduced body schema availability. This paper calls this condition a self-location-dominant state. In this state, viewpoint-directed agency is retained, whereas environment-directed agency and body ownership are constrained. Nevertheless, events such as viewpoint motion, impact, contact, and direct address may be experienced not merely as changes within an image, but as events concerning the viewpoint position at which the self is located. This paper examines this structure by connecting research on presence, the sense of embodiment, bodily self-consciousness, and the minimal self. The minimal self in immersive video is thereby redescribed not primarily in terms of agency or ownership, but in terms of viewpoint-based self-location established under conditions in which the contribution of the body schema is reduced. This perspective provides a basis for theorising self-experience in non-interactive immersive media and for reconsidering the relation between body, viewpoint, and presence in XR.

## I. INTRODUCTION: Reconsidering the Self in XR

What, then, is the self? Since Descartes, this question has repeatedly been posed as one of the fundamental problems of philosophy. Following the grounding of the self as a thinking subject in the dictum "*I think, therefore I am*", the phenomenological tradition reformulated the self as a being that relates to the world through the body. In other words, the self came to be understood not as a merely internal consciousness, but as a being situated in the world, perceiving and acting through bodily mediation.

Yet immersive media environments within XR (Extended Reality), and immersive video in particular, compel a reconsideration of this body-mediated understanding of the self. The present paper examines immersive video, understood here as a format that presents pre-recorded 180-degree or 360-degree footage through high-resolution head-mounted displays. It differs both from interactive VR experiences in which the user participates in a virtual space through an avatar, and from conventional two-dimensional video viewing on a screen. Immersive video can evoke a subjective experience involving bodily and spatial qualities, yet the video itself remains pre-recorded and non-interactive, and the experiencer cannot actively change the presented environment. In this respect, immersive video constitutes a boundary form of XR media situated between interactive VR and conventional two-dimensional video viewing.

The sense of being there, or presence, has been treated as a central issue in XR experience. In previous research, the technical conditions on the side of the system that enable immersion have been distinguished from presence, the subjective sense of being there that arises under those conditions (Slater, 2009; Skarbez et al., 2017a). Slater (2009) distinguished two illusions constitutive of presence: Place Illusion (PI) and Plausibility Illusion (Psi). PI refers to the illusion of being in the place depicted by a virtual environment, even though the experiencer knows that they are physically located elsewhere. In Slater's account, however, PI is not merely an abstract sense of spatial location; it is supported by sensorimotor contingencies through which perception is updated in response to bodily movement, such as head movement, gaze shift, postural change, approach, or avoidance. Psi, by contrast, refers to the illusion that events and objects within the environment are coherent and that what is occurring there is actually taking place (Slater et al., 2022; Skarbez et al., 2017b).

Immersive video occupies a distinctive position in considering the conditions under which presence is established. Because the user can alter the field of view through head rotation, immersive video does not entirely lack sensorimotor contingencies. Such contingencies, however, are largely restricted to changes in gaze direction and head orientation. The user cannot change the recorded environment through walking, approaching, avoiding, grasping, or manipulating. In many cases, no explicit body or avatar corresponding to the user is provided. Presence in immersive video should therefore not be simply equated with the full PI assumed by Slater. Rather, it should be understood as presence that emerges under conditions in which limited sensorimotor contingencies, namely the relation between head rotation and visual updating, are retained, while active intervention in the environment and bodily possibilities for action are substantially constrained.

This point produces an important tension between research on presence and research on embodiment. Slater's PI is not identical with body ownership, yet the concept is not wholly separable from the relation between bodily movement and perceptual updating. By contrast, research on bodily experience in VR has described the sense of embodiment as comprising three components: self-location, agency, and body ownership (Kilteni et al., 2012). In Kilteni et al.'s framework, self-location is not presence itself, but concerns where the self is located in relation to the body. Accordingly, research on presence and research on embodiment both involve self-location, but position it within different theoretical relations. In Section II, the present paper clarifies this relation through body schema, body ownership, agency, and bodily self-consciousness.

In the present paper, presence is treated as a broad subjective state in which the user feels that they are in the presented environment. Self-location, by contrast, is closely related to presence, but is not identical with presence itself; it refers more specifically to the problem of where the self is situated in space. The argument of this paper is therefore not that presence should be reduced to self-location. Rather, the issue is that, in immersive video, self-location becomes relatively dominant over agency and ownership within the self-experience that accompanies presence.

At this point, a careful distinction within agency is required. Agency is not completely lost in immersive video. The user can move the head, select the direction of gaze, and explore the surrounding visual field. In this sense, a limited form of agency related to the orientation of the viewpoint is retained. This agency, however, is not *environment-directed agency*, by which the user acts upon the recorded environment and changes objects or events within it. What is constrained in immersive video is agency as an action that changes the environment; the minimal sensorimotor contingency involved in selecting the viewpoint and updating the field of view is not entirely absent. Introducing this distinction allows presence in immersive video to be reconsidered not as presence without agency, but as presence supported by limited viewpoint-related contingency while lacking environment-directed agency.

This structure unsettles assumptions about embodiment in VR research. The user may experience a strong sense of spatial presence, yet this experience is not sufficiently grounded in bodily movement, manipulation, or body ownership over an avatar. A viewpoint exists, but the body corresponding to that viewpoint is not explicitly given. The self is experienced as located in space, even though no explicit bodily model corresponding to that location is stably constituted. The absence of the body in this context, however, should not be understood as the loss of the physiological body or postural control. The user's body is, in fact, maintained on a chair, and head movement, eye movement, postural adjustment, vestibular sensation, and proprioception remain operative. The issue is the extent to which bodily posture, bodily outline, centre of gravity, and possibilities for movement are used as explicit and stable cues when self-location is constituted within the mediated visual space. The present paper therefore understands this state not as the disappearance of the body schema, but as reduced *body schema availability* in mediated self-location.

From this perspective, the absence of the body in immersive video is not merely a technical deficiency. Rather, it is a media-specific feature that alters the conditions under which self-experience is constituted. In VR where an avatar or virtual body is provided, the central issue is how body ownership and agency are established. In immersive video, by contrast, presence is not reinforced by providing a body. Rather, self-location emerges through the viewpoint while the body is not explicitly given. In this sense, immersive video should be understood not as a medium that enhances bodily self-consciousness as a whole, but as a medium that selectively redistributes its components.

The present paper focuses precisely on this redistribution. In immersive video, the user is not a subject who acts upon the recorded environment, nor does the user stably own a body corresponding to the viewpoint. Nevertheless, the user may often experience the self as being located at a point within the video space. This self-experience differs from merely third-personal observation in conventional two-dimensional video viewing. In third-personal observation, changes in viewpoint or camera shake are typically processed as events within the image. In immersive video, by contrast, when the stability of the viewpoint breaks down, the body schema that had been backgrounded may return to the foreground in an incomplete form. At such moments, camera shake or impact may be experienced not merely as movement on a screen, but as the destabilisation of the very viewpoint position at which the self is located.

The present paper understands subjective experience in immersive video not as bodily extension, but as spatial self-location that emerges under conditions in which the contribution of the body schema is reduced. Spatial self-location here refers to an experience in which the self is located at a point in space without explicitly and stably relying on bodily motor possibilities, posture, touch, vestibular sensation, or proprioception. This is not a self entirely without a body. Rather, it is a state in which, despite the continued presence of the physiological body, body schema availability is reduced in self-location within the video space, and viewpoint-based self-location becomes relatively dominant. This paper calls this condition a *self-location-dominant state.*

The aims of the present paper are threefold. First, it formulates immersive video as a non-interactive immersive medium and as a theoretical boundary case that reveals the dissociability of body schema and self-location. Second, it shows the limitations of explaining the minimal self solely in terms of agency and ownership, and positions self-location as an independent analytic axis. Third, through phenomenological observations of immersive video, it clarifies how the body schema is backgrounded and incompletely returns across ordinary viewing, viewpoint disruption, and re-stabilisation.

The argument of the present paper is not that the body should be excluded from presence. On the contrary, immersive video reveals, through the reduced contribution of the body schema, the extent to which the body schema ordinarily supports self-experience. The body does not always appear as an explicit object. In many cases, it operates transparently as a background condition supporting self-location. When this transparency is disturbed by camera shake, impact, contact, or direct address, the body returns not as a complete body image, but as incomplete bodily awareness. This incomplete return of the body is the clue to the distinctive character of the minimal self in immersive video.

## II. THEORETICAL BACKGROUND: Body Schema, Ownership, and Self-Location

The tradition of understanding the self in relation to the body has been strongly developed in phenomenology. In Husserlian phenomenology, consciousness is understood as always having intentionality, as always being directed towards something, and the self is understood in relation to the world (Husserl, 1907). Heidegger developed this line of thought as *being-in-the-world*, showing that human existence is always constituted through engagement with the world (Heidegger, 1927). Merleau-Ponty further situated the body not as a merely physical object, but as the lived body (Merleau-Ponty, 1945, 1962). For Merleau-Ponty, the body is not an object located in the external world; rather, it is the medium through which the world is perceived and engaged. This view is grounded in an account of behaviour not as mere reflex or mechanical response, but as a structure organised through the relation between body and environment (Merleau-Ponty, 1942, 1963). In this trajectory, self-understanding was displaced from the mind-body dualism expressed in Descartes' "*I think, therefore I am"* (Descartes, 1637) towards a bodily understanding of the self that may be summarised as "*I can, therefore I am*".

This bodily possibility concerns not only the body image that is consciously objectified, but also the body schema that pre-reflectively supports action and perception. Gallagher (2005) distinguishes the body image, as a reflective or conscious bodily representation consisting of perceptions, attitudes, and beliefs about one's own body, from the body schema, as a system of pre-reflective sensorimotor capacities that supports posture, motor control, and the spatial organisation of action without conscious monitoring. What matters for the present paper is not whether the user explicitly sees their own body, but to what extent the body schema is mobilised when self-location is established in immersive video. In VR research as well, the sense of embodiment has been treated as a multidimensional construct that includes body ownership, agency, and self-location, and its assessment methods and the relations among its components have been systematically reviewed (Kilteni et al., 2012; Guy et al., 2023). This supports the standpoint of the present paper, which analyses self-experience in immersive video not through a single measure of presence, but through the relation among body ownership, agency, body schema, and self-location.

In this context, research on bodily self-consciousness is particularly important. Blanke and Metzinger (2009) examined minimal phenomenal selfhood through the interrelation of bodily identification, self-location, and first-person perspective, and argued that self-experience should be understood in terms of the global and unitary aspects of bodily self-consciousness, especially the relations among the body, spatial location, and first-person perspective. What matters here is that the minimal conditions of self-experience are not necessarily reducible to an explicit body image or active action. Rather, bodily self-consciousness is constituted by several components: where the self is located, with which body it is identified, and from which viewpoint the world is given. This framework is important for understanding self-experience in immersive video, because immersive video provides a first-personal visual field and spatial surround, while bodily identification and active action towards the environment remain substantially constrained.

Gallagher (2000) discussed the minimal self as a pre-reflective form of self-experience prior to the narrative self, and introduced the distinction between the sense of agency and the sense of ownership in order to analyse its experiential aspects. When one moves one's arm, one feels that "I am moving it"; in this case, agency is established. At the same time, when the moving arm is felt not as someone else's but as one's own body, ownership is established. The strength of Gallagher's account lies in situating the self not as an abstract consciousness, but within the coherence of bodily movement and sensory feedback.

The vocabulary that explains the minimal self primarily in terms of agency and ownership functions most strongly in situations in which the body schema is relatively stable and available. That is, this vocabulary has strong explanatory power in situations where active movement is possible, where motor commands correspond to sensory feedback, and where self-generated movement is experienced on the basis of that coherence. Classically, it has been shown that the perceived times of intentional actions and their sensory consequences are drawn towards one another in subjective time, and this intentional binding has been widely used as an implicit measure of the sense of agency (Haggard et al., 2002; Moore & Obhi, 2012). Subsequent research, however, has indicated that intentional binding is not a pure measure of agency, but can also vary depending on causal relations between action and outcome, prediction, postdictive inference, belief, and the integration of multiple agency cues (Buehner & Humphreys, 2009; Buehner, 2012; Synofzik et al., 2008). The present paper therefore positions intentional binding not as direct evidence of agency itself, but as a classical finding indicating that the relation between active action and sensory outcome is involved in the formation of self-experience.

In addition, the distinction between voluntary movement and externally induced bodily events is related to the conscious processing of bodily events (Tsakiris & Haggard, 2003). Furthermore, the subjective simultaneity between self-generated movement and auditory feedback can be recalibrated according to the range of experienced delays (Toida et al., 2014). The author has previously examined the possibility that the temporal integration of motor commands and sensory feedback is related to self-generated experience and agency-related processing (Toida et al., 2014, 2016). These findings indicate that self-experience is supported by the relation between action and sensory outcome. In immersive video, however, this action–sensory feedback loop is substantially constrained. The user can look around, and thus a limited form of agency remains in the selection of gaze and head direction. Yet the user cannot intervene in the recorded environment through walking, approaching, avoiding, grasping, or manipulating, and no correspondence is established between motor commands and sensory feedback that changes the recorded environment itself. Accordingly, if presence or self-location arises in immersive video, the corresponding self-experience cannot be fully explained by a model centred on environment-directed agency.

Body ownership, by contrast, is also not fixed. The Rubber Hand Illusion shows that, through the integration of visual, tactile, and proprioceptive information, a rubber hand that is not part of the biological body can be felt as one's own hand (Botvinick & Cohen, 1998). This finding indicates that ownership is not an intrinsic property of the body itself, but a plastic experience constituted through multisensory integration. Subsequent research has further shown that the integration of visual, tactile, and proprioceptive information in the Rubber Hand Illusion can be explained within a framework of Bayesian causal inference; body ownership can thus be understood as an experience that depends on the estimation of whether multiple body-related cues originate from the same body (Samad et al., 2015). In addition, after acquired limb amputation due to accident or other causes, phantom limb phenomena have been reported, in which pain or a sense of position arises in a limb that no longer exists (Ramachandran & Hirstein, 1998). This indicates that body image and bodily position sense do not simply depend on the actual existence of the biological body, but may be reconfigured through central representations and neural plasticity.

Furthermore, self-location is also not fixed to the position of the biological body. Research on out-of-body experiences has reported that electrical stimulation of temporo-parietal regions, including the right angular gyrus, can induce out-of-body-like experiences, illusory experiences of body parts, and vestibular sensations of whole-body displacement (Blanke et al., 2002, 2005). Research on bodily self-consciousness using VR has also shown that, through conflict between visual and somatosensory information, participants can feel a virtual body as their own body and mislocalise self-location from the position of the biological body towards the virtual body (Lenggenhager et al., 2007). These findings indicate that the self is not always located within the biological body, and that self-location itself can be reorganised depending on the integrative relation among body schema, vestibular sensation, proprioception, visual information, and first-person perspective.

This point is further supported by recent research on visual–vestibular integration. Wu et al. (2024) combined mixed reality with a motion platform and showed that an out-of-body-like illusion can be induced by manipulating visual–vestibular integration. More specifically, congruent visual–vestibular stimulation in a self-centred reference frame has been reported to produce an OBE-like illusion involving elevated self-location, disembodiment, and lightness. This study does not directly concern immersive video itself. Nevertheless, it is importantly connected to the problem addressed in this paper insofar as it shows that the integrative relation between visual information and vestibular sensation is involved in self-location and experiences of disembodiment. Immersive video does not involve artificial vestibular stimulation or a motion platform. Yet, through wide-field video and visual updating in response to head movement, it presents the viewpoint in video space as a primary cue for self-location. Immersive video should therefore be understood as distinct from experimental systems that directly manipulate visual–vestibular integration, but as a complementary boundary case for examining how self-location is stabilised, and how it breaks down, within the relation between visual frames of reference and bodily sensation.

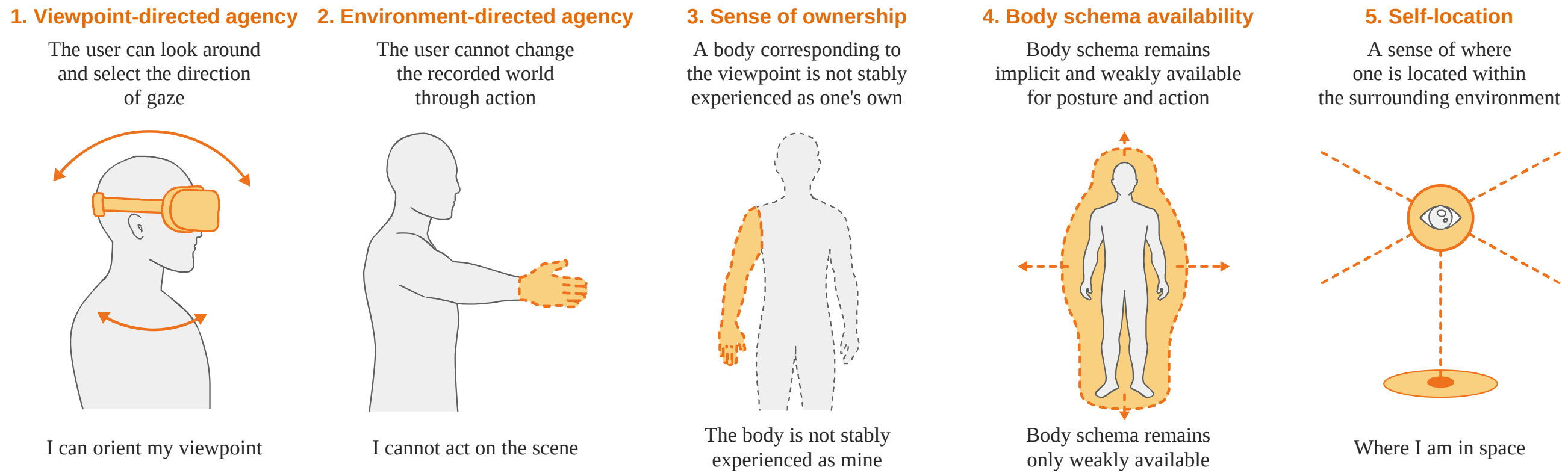

**FIGURE 1** Core components of self-experience in immersive video. The diagram distinguishes viewpoint-directed agency, environment-directed agency, sense of ownership, body schema availability, and self-location.

The relation between presence and bodily self-consciousness must also be treated carefully. Herbelin et al. (2015) present a framework for examining the experience of presence in VR in relation to the neural mechanisms of bodily self-consciousness. From this perspective, presence should be examined not only in terms of environmental realism or device immersion, but also in relation to the components of bodily self-consciousness: where the self is located, how the body corresponding to that location is identified, and from which viewpoint the world is given. The argument of the present paper, however, is not that presence should be reduced to bodily self-consciousness. Rather, the distinctive character of immersive video lies in the fact that the components of bodily self-consciousness are not enhanced as a whole, but are selectively retained or reduced. That is, viewpoint and self-location are relatively strongly established, while body ownership, body schema availability, and environment-directed agency are constrained.

On this basis, agency, ownership, body schema, and self-location are normally interrelated, but are in principle dissociable. Indeed, research on VR environments has also described the sense of agency and the sense of body ownership as dynamic components that can be selectively altered by different manipulations, while also interacting depending on context (Girondini et al., 2025). Immersive video is a site in which this dissociability appears as a concrete media experience. That is, there may be cases in which environment-directed agency is constrained, ownership is unstable, and body schema availability is reduced, while self-location is maintained relatively strongly (Fig. 1).

Taken together, this theoretical background suggests that self-experience in immersive video should not be treated merely as a form of video reception in which the body is absent, but as a state in which the components of bodily self-consciousness are redistributed. The self established here is neither a subject who actively intervenes in the environment, nor a subject who stably owns a body corresponding to the viewpoint. Rather, it is a self located in space under conditions in which the explicit contribution of the body schema is reduced, supported by limited viewpoint operation and wide-field video. The following section examines, through phenomenological case analysis, how this asymmetry appears in the viewing experience of immersive video.

## III. CASE ANALYSIS: A Phenomenology of Self-Location in Immersive Video

In order to clarify the distinctive character of self-location in immersive video, it is first necessary to consider how existing moving-image media have constituted the position of the viewer. In traditional moving-image reception, the relation between the viewer and the video world has primarily been described through narrative personhood and the configuration of viewpoint (Branigan, 1984; Bordwell & Thompson, 2010; Sobchack, 1992). In film theory, the POV shot functions as a device that allows the viewer to share the visual field of a character, thereby producing a quasi-first-personal experience (Branigan, 1984). Direct address, in which a character looks into the camera and speaks directly to the viewer, produces a second-personal experience in which the viewer is positioned as “you” (Gibbons & Whiteley, 2021). Third-personal experience refers to the classical mode of viewing in which the viewer observes events from outside, and much of standard film grammar is grounded in this observational configuration (Bordwell & Thompson, 2010).

Personhood and viewpoint configuration, however, describe how something is seen; they do not sufficiently explain how the body participates in self-experience. A first-person perspective does not necessarily entail the stable mobilisation of the body schema. A second-personal address does not necessarily mean that one’s body is constituted within that space. Even when the experience appears to be third-personal observation, camera shake or impact can bring the viewpoint position to the foreground as the position of the self. Accordingly, while the present paper uses personhood as an introductory descriptive framework, the centre of analysis is shifted to the relation between body schema and self-location.*

This issue becomes particularly salient when the stability of the camera breaks down. In 360-degree video, camera stability has been shown to affect cybersickness, and in VR more generally, mismatch between visual and vestibular information is related to bodily discomfort and alterations in self-motion perception (Litleskare & Calogiuri, 2019; Nürnberger et al., 2021; Chang et al., 2020). Moreover, in presence research, disruptions in engagement with an immersive environment have been discussed as breaks in presence (Slater & Steed, 2000; Brogni et al., 2003). In this sense, camera shake and impact in immersive video can be understood not merely as movement within the image, but as a media-specific break in presence that temporarily renders explicit the relation between visual self-location and the physiological body.

The following analysis draws on phenomenological observations made by the author while experiencing several Apple Immersive Video works using Apple Vision Pro. The descriptions in this section do not constitute empirical proof about immersive video in general. Rather, they function as phenomenological vignettes intended to extract experiential structures that are difficult to capture within existing research on presence and embodiment. The aim here is not to make statistical claims about immersive video in general, but to clarify conceptual distinctions that may be tested in subsequent empirical research. The focus is not the content of the videos themselves, but how self-location and bodily awareness change between states in which the viewpoint is stable and states in which the camera is shaken or impacted.

During ordinary viewing, the experiencer of immersive video often obtains a strong sense of presence, as if being within the video space. When experiencing, for example, live performance footage or sports footage, the experiencer does not merely watch events on a screen, but feels surrounded by the video space and situated at a particular viewpoint within it. Yet this being there is not constituted together with one's own arms, legs, posture, centre of gravity, or possibilities for movement. Rather, only the viewpoint position appears to be fixed within the space, while the body schema remains backgrounded. What is established here is not a first-personal bodily experience accompanied by body ownership, but a state in which self-location is established through the viewpoint while the explicit contribution of the body schema is reduced.

This structure becomes especially salient in scenes involving camera shake or impact. In *2024 NBA All-Star Weekend* (Apple, 2024a), the camera is attached to a basketball goalpost. At the moment when a player completes a dunk, the entire goalpost vibrates and the image shakes substantially. At this moment, the author's experience was not merely one of externally observing the game. Rather, self-location seemed to be temporarily drawn towards the viewpoint attached to the goal. Yet, at the same time, no body image corresponding to that position emerged. In other words, the self was located at that viewpoint, but the body schema supporting that self was not clearly constituted.

This experience differs from the simple explanation that "the camera becomes the body". If the camera were stably owned as a body, the vibration of the goalpost should be integrated as a continuous transformation of body ownership. What actually arises, however, is not the feeling of owning a body, but a discomfort in which the absence of the body is temporarily exposed through the destabilisation of the viewpoint. In other words, camera shake does not strengthen body ownership, but reveals the incompleteness of the bodily basis of self-location that had previously been smoothly established.

A similar phenomenon can be observed in *Man vs. Beast* (Apple, 2025a), a documentary work on rodeo following professional bull riders, when a bull collides with the camera. The experiencer cannot exercise environment-directed agency, such as avoiding or defending against the bull. At this moment, the event is experienced not as something occurring at a distance within the image, but as if an external force had been received at the camera position. This discomfort, however, is not a stable sense that "my body has been struck". Rather, it is an incomplete return of bodily awareness: the body is recalled, yet its posture, outline, centre of gravity, and possibilities for movement cannot be clearly constituted. Here, self-location is strongly tied to the viewpoint position, while the body corresponding to that viewpoint is not clearly formed.

The same structure appears in *Metallica: Apple Immersive* (Apple, 2025b), when a balloon touches the camera during the live performance. Although the contact occurs with the camera, it is experienced not merely as contact within the image, but as an event at the viewpoint position. Yet here again, the camera is not stably owned as a body. Rather, self-location is first established with very little explicit mobilisation of the body schema, and the contact exposes the fact that this self-location is not sufficiently supported by the body schema. In this sense, contact does not constitute body ownership, but paradoxically reveals that body ownership has not been established.

By contrast, in *Alicia Keys: Rehearsal Room* (Apple, 2024b) and *Metallica: Apple Immersive*, when performers sing towards the camera, the experiencer temporarily obtains

* Similar techniques exist in two-dimensional moving images, where camera shake or rapid movement can induce bodily responses in viewers. For example, in found-footage films and films using subjective camera perspectives, camera instability is experienced in relation to the body of the camera-holder or the character within the film. In music videos as well, camera shake and changes in speed can generate bodily rhythm or a sense of movement in the viewer when synchronised with dance, beat, or editing. In these two-dimensional images, however, such movement is often received as belonging to the body of a character, the movement of a camera operator, or the expressive intensity of the image. In immersive video, by contrast, the entire field of view presented in the head-mounted display is displayed as surrounding space. Camera shake and impact may therefore be experienced not merely as movement on a screen, but as the destabilisation of the very viewpoint at which the self is located.

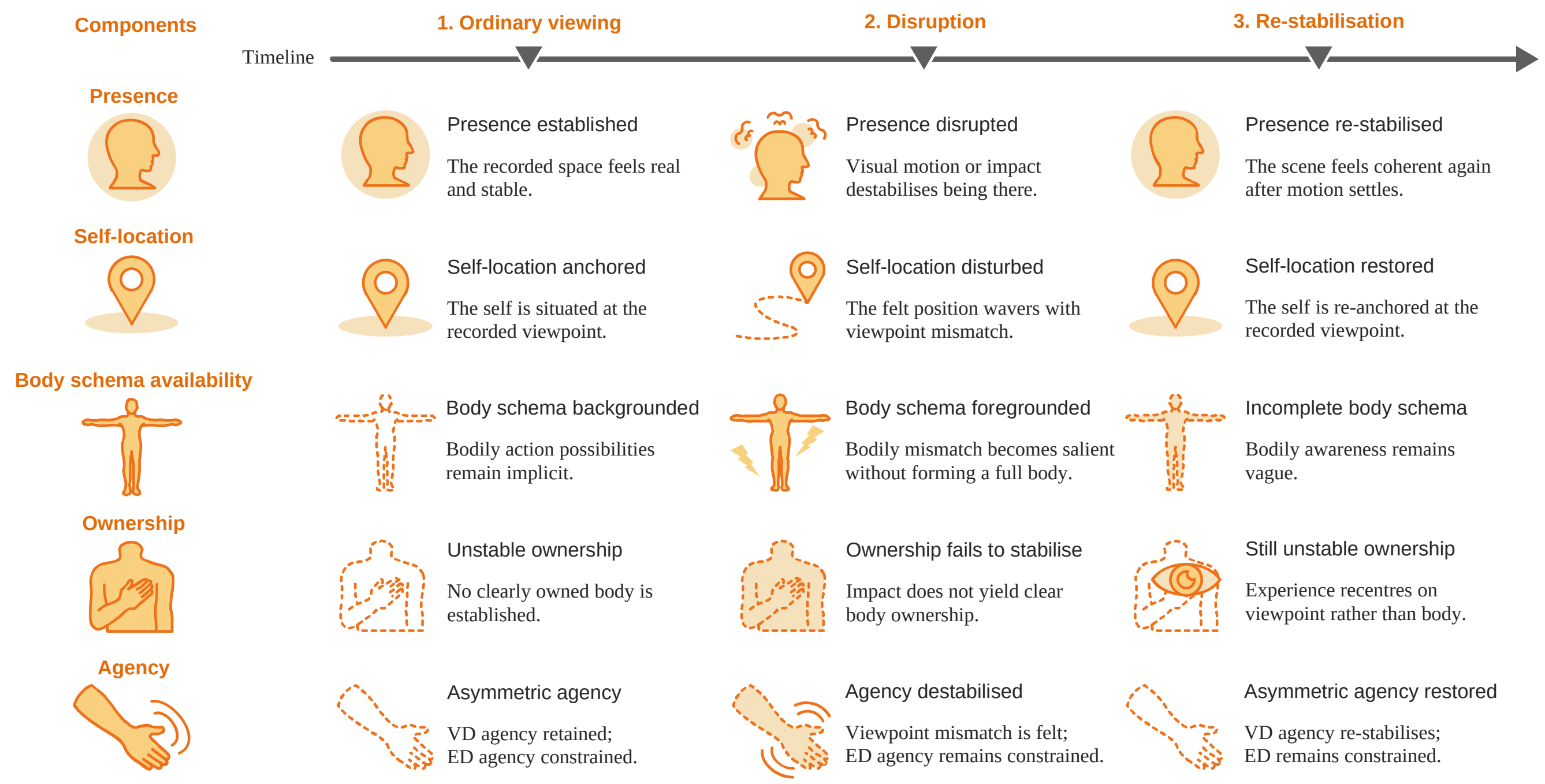

**FIGURE 2** Temporal dynamics of self-experience in immersive video across ordinary viewing, disruption, and re-stabilisation. VD and ED denote viewpoint-directed agency and environment-directed agency, respectively.

a second-personal experience of being addressed as “you”. This address, however, does not directly lead to the stable constitution of the body schema. In ordinary face-to-face interaction, the self who is addressed is situated in the world as an embodied interlocutor. In immersive video, by contrast, the addressee of the address is not so much an embodied interlocutor as a viewpoint fixed within space. What arises here is not interaction with another embodied person, but a distinctive self-experience in which a bodiless viewpoint is treated as “you”.

These cases show that self-experience in immersive video cannot be adequately captured by the film-theoretical classification into first-, second-, and third-personal experience. At the moment of camera shake, first-personal self-location arises. Yet it is not a first-personal experience in which one enters the space with one’s own body. At the moment when a performer addresses the camera, second-personal address arises. Yet it is not a face-to-face relation accompanied by the body. During ordinary viewing, a structure close to third-personal observation also remains. Yet when the camera is impacted or touched, this observational distance temporarily breaks down, and the viewpoint position comes to the foreground as the position of the self.

The camera in immersive video should therefore be understood not as a proxy for the body, but as a locating device that induces self-location. The camera gives the viewpoint from which the experiencer sees the world. Yet it does not necessarily give a body that is owned. This is a crucial feature of immersive video. A viewpoint is given, but a body is not. Self-location is established, but the body schema supporting that self-location is not explicitly constituted. Camera shake and impact temporarily expose this asymmetry.

Understood in this way, the ordinary state of immersive video should be understood not as the enhancement of body ownership, but as the transparency of the body schema. The user is not simply forgetting the body. More precisely, in the constitution of spatial self-location, the body schema is not explicitly or stably mobilised. This state is smoothly maintained as long as the viewpoint remains stable. When the stability of the viewpoint is disturbed by camera shake, impact, contact, or direct address, however, bodily awareness that had been backgrounded comes to the foreground in an incomplete form. The body is not recovered; it is incompletely reactivated.

This “incomplete return of the body” indicates the distinctive character of the minimal self in immersive video. During ordinary viewing, presence and self-location are established, while the body schema is transparent. When the viewpoint breaks down, presence and self-location are temporarily destabilised, and bodily awareness comes to the foreground. Yet the body is not recovered as a body image with clear posture, outline, centre of gravity, or possibilities for movement. Rather, the incompleteness of the body is experienced precisely at the moment when the need for a body is exposed, yet that body cannot be sufficiently constituted. This process provides a phenomenological clue that, in immersive video, self-location is established in relative dissociation from the body schema (Fig. 2).

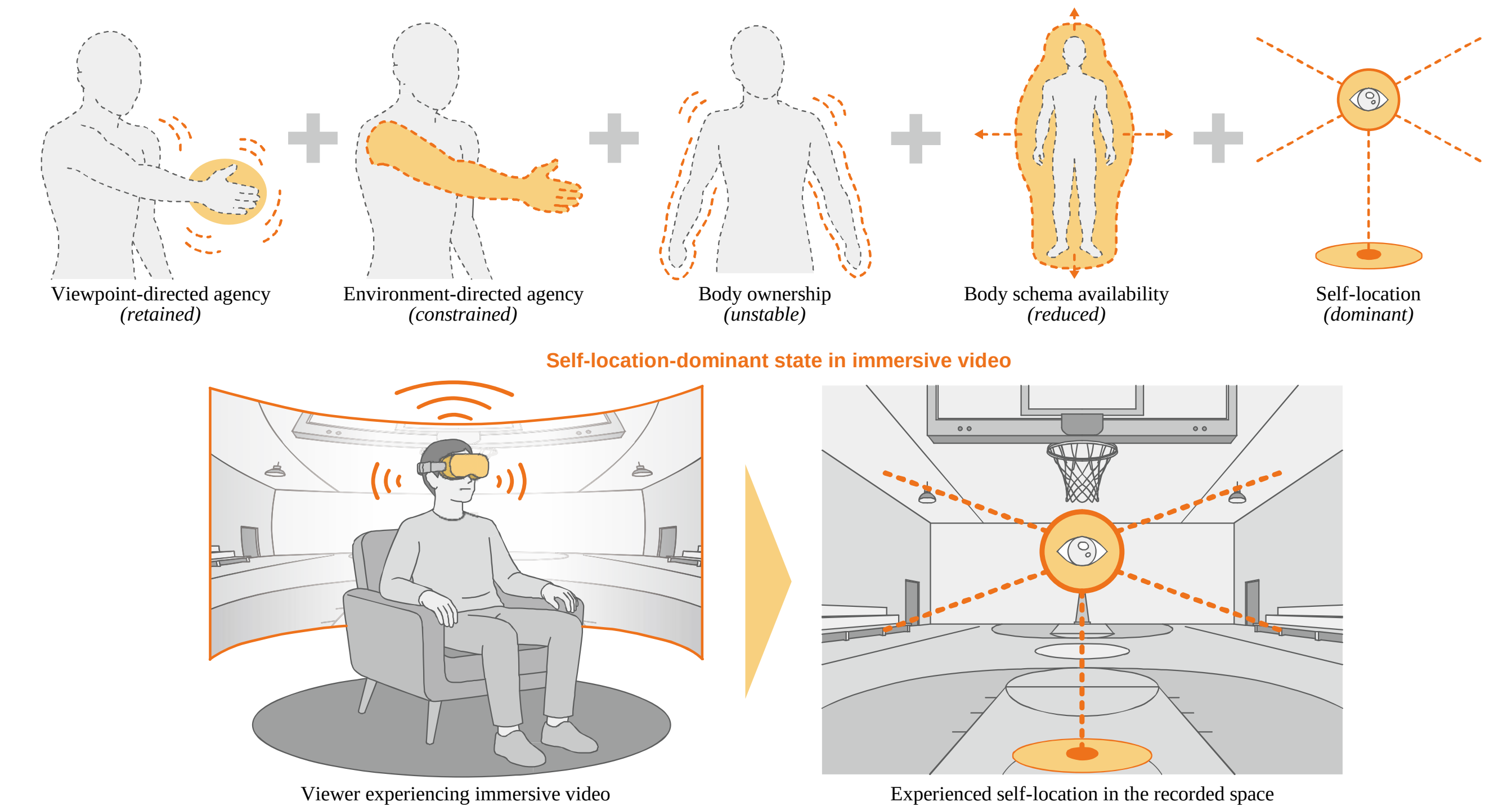

**FIGURE 3** Reconfiguration of self-experience into a self-location-dominant state in immersive video. The diagram shows how retained viewpoint-directed agency, constrained environment-directed agency, unstable body ownership, reduced body schema availability, and dominant self-location jointly constitute the state.

## IV. DISCUSSION: Self-Location-Dominant State

Why, then, can the self be located in space in immersive video despite the insufficient establishment of bodily action and stable body ownership? As shown in Section II, agency, ownership, body schema, and self-location are normally interrelated, yet are in principle dissociable. As shown in Section III, in immersive video, self-location temporarily comes to the foreground through events such as viewpoint shake, impact, contact, and direct address, while the body schema corresponding to that position is not stably constituted. The present paper calls this condition a self-location-dominant state: a state in which the self is located at a viewpoint in space despite constrained environment-directed agency, unstable body ownership, and relatively reduced body schema availability. This is not a state in which the body schema has disappeared. Rather, it is a state in which self-location becomes relatively dominant over agency, ownership, and body schema in the constitution of the minimal self.

The significance of this concept lies in understanding immersive video not as a weak form of embodiment in VR, but as a media condition in which the components of bodily self-consciousness are redistributed. Previous VR research has primarily discussed how body ownership and agency are established through avatars, visuo-tactile synchrony, and active manipulation. By contrast, in immersive video, no body or avatar is explicitly given, and the possibility of acting upon the environment is restricted. What is at stake, therefore, is not how body ownership is enhanced, but why the self can nevertheless be located in space when body ownership and active intervention in the environment are insufficiently established.

What is important in considering this question is not to treat agency as a monolithic concept. Agency is not completely lost in immersive video. The user can rotate the head, select the direction of gaze, and explore the presented space. In this sense, *viewpoint-directed agency*, or a limited sensorimotor contingency related to the viewpoint, is retained. Yet the user cannot change the recorded environment through walking, approaching, avoiding, grasping, or manipulating. Environment-directed agency is therefore substantially constrained. Presence and self-location in immersive video should thus be understood not as arising under the complete absence of agency, but under an asymmetric condition in which viewpoint-directed agency remains while environment-directed agency is constrained.

This account does not reject Gallagher's framework of the minimal self. Rather, it clarifies the conditions under which agency and ownership function effectively as analytic axes of the minimal self. In situations where there is bodily movement, motor intention, sensory feedback, and the integration of these factors, agency and ownership are central for explaining pre-reflective self-experience. In a situation such as immersive video, however, where no correspondence exists between motor commands and environmental change and where the body corresponding to the viewpoint is not stably owned, self-experience takes a different configuration.

In this case, what becomes central is not agency as a subject acting upon the environment, but viewpoint-based self-location.

This temporal process was presented in Fig. 2. Fig. 3 further organises the same structure as a reconfiguration of self-experience into a self-location-dominant state. During ordinary viewing, presence and self-location are relatively stably established, while the contribution of the body schema is backgrounded, and body ownership and environment-directed agency remain limited. Viewpoint-directed agency, however, is retained. When camera shake, impact, or contact renders mismatch between visual information and vestibular or proprioceptive sensation explicit, presence and self-location are temporarily destabilised, and the body schema comes to the foreground in an incomplete form. This is not, however, the recovery of stable body ownership, but an incomplete return of bodily awareness in which the body image cannot be clearly reconstituted.

On this basis, the minimal self in immersive video should be understood neither as a subject that acts upon the environment through agency, nor as a subject that stably owns a body through ownership. Rather, it is a self located at a viewpoint in space under conditions of reduced body schema contribution. More precisely, self-experience in immersive video is constituted as an asymmetric combination of viewpoint-based self-location, retained viewpoint-directed agency, unstable body ownership, reduced body schema availability, and constrained environment-directed agency.

Several objections may be raised against this argument. The first is that the experience described here is merely third-personal observation. If it were processed solely as third-personal observation, however, camera shake or impact would be experienced primarily as an external change within the image. In immersive video, by contrast, camera shake or impact may be experienced as a change concerning the position of the self. This indicates that the experience in question is not mere observation, but includes self-location.

The second objection is that the camera simply functions as a proxy body. The camera, however, is not stably owned as a body. If the camera were experienced as a body, shake or impact would be integrated as a continuous change in body ownership. What actually arises is not stable body ownership, but a sudden return of bodily awareness accompanied by discomfort. In this respect, the camera should be understood not as a substitute body, but as a locating device that induces self-location without the stable constitution of the body schema.

The third objection is that the body has not disappeared, but merely remains latent. This objection is partly correct. During the experience of immersive video, the user's biological body is in fact often maintained on a chair, and head movements, eye movements, postural adjustment, vestibular sensation, and proprioception remain operative. The issue addressed by the present paper, however, is not whether the physiological body exists, but the extent to which the body schema contributes to the formation of self-location. In immersive video, during ordinary viewing, the body schema is less likely to be used as an explicit and stable cue for spatial self-location. The fact that the body is recalled during disruption indicates, rather, that in ordinary viewing the body schema is backgrounded in the constitution of self-location.

The fourth objection is that presence itself presupposes the body. This objection is important, and Slater's (2009) Place Illusion cannot simply be invoked as a basis for "bodiless self-location". In Slater's account, PI is not body ownership itself, but it is supported by sensorimotor contingencies through which perception is updated in response to head movement and bodily movement. Presence in immersive video should therefore not be equated with the full PI assumed by Slater. It should instead be understood as presence that emerges under conditions in which some requirements of PI are met by head rotation and wide-field video, while action-contingencies through bodily movement or environmental manipulation are substantially restricted. The argument of the present paper is not that embodiment should be removed from Slater's PI. Rather, it focuses on the fact that immersive video can nevertheless generate an experience close to self-location even under conditions in which the sensorimotor contingencies supporting PI are limited. In this sense, the present paper does not use Slater's PI concept as its direct foundation, but takes as its main axis the framework distinguished by Kilteni et al. (2012), namely self-location, agency, and body ownership, in order to discuss the distinctive form of self-location in immersive video.

A fifth objection is that presence in immersive video is not essentially different from observational presence in theatre or cinema. Certainly, there are situations in which experiencers feel present in a place even without actively intervening in the environment. Theatre, cinema, and live performance are typical examples. The emergence of presence in a non-interactive situation is therefore not, in itself, exceptional. The point of immersive video, however, is not merely that presence arises despite non-interactivity. What matters is that this presence is established not as observation from outside a screen, but as viewpoint-based self-location within a surrounding space presented inside a head-mounted display. In immersive video, an observational structure remains, yet the viewpoint is organised as the position of the self. In this respect, immersive video should be understood not as mere spectatorship, but as a boundary experience in which spectatorship and self-location overlap.

This argument can also be supported from the perspective of multisensory integration. In classical Bayesian models of sensory integration, perception is constituted through the integration of multiple sensory cues, and the weighting of each cue varies according to its reliability, namely the smallness of its estimated variance (Ernst & Bülthoff, 2004). Subsequent research, however, has described multisensory integration not merely as reliability weighting, but as Bayesian causal inference that estimates whether multiple sensory cues originate from the same external cause (Körding et al., 2007; Shams & Beierholm, 2010). Within this framework, in immersive video, the wide field of view, synchrony with head rotation, and spatial coherence allow the video space to be estimated as the principal frame of reference for self-location. Conversely, when camera shake or impact renders mismatches between visual information and vestibular or proprioceptive information explicit, the estimation that these cues arise from the same cause becomes unstable, and the body schema that had been backgrounded comes to the foreground. The self-location-dominant state in immersive video should therefore be understood not as simple visual dominance, but as a reconfiguration of reliability weighting and causal inference among visual information, vestibular sensation, proprioception, and body schema availability.

This shift in the frame of reference towards the video space is also related to the weakening of the visual frame of reference provided by the physical environment through wide-field video, and to the induction of vection through optic flow. An independent visual frame of reference has been shown to reduce simulator sickness (Duh et al., 2004), and in VR research, static and dynamic rest frames have been examined as possible means of reducing visually induced discomfort (Cao et al., 2018). Optic flow can also induce visually mediated self-motion perception, that is, vection, indicating that visual information may be strongly involved in self-motion perception and spatial location (Keshavarz et al., 2015). Thus, in immersive video, proprioception and vestibular sensation do not disappear; rather, during ordinary viewing, these cues are relatively backgrounded in the constitution of self-location, and the visual frame of reference within the presented video space functions as the principal cue for self-location.

This shift in the frame of reference towards the video space is not, however, always stable. When camera shake or impact renders mismatch between visual information and vestibular or proprioceptive information explicit, the body schema that had been backgrounded once again appears in the foreground of self-experience. Here lies the tension between presence and the body in immersive video. The argument is not that stronger presence necessarily entails a stronger foregrounding of the body. Rather, when presence is smoothly established, the body schema is transparent. The body is not always foregrounded as the basis of presence; instead, it returns, especially in an incomplete form, when presence breaks down. In this sense, the body in immersive video is not a body that is stably owned, but a body that returns incompletely at the moment of disruption.

This argument has two implications for XR research. First, research on presence must address not only immersion or realism, but also the relation between self-location and body schema availability. Second, research on the self in XR must address not only the provision of a body, as in avatar embodiment or full-body illusion, but also the conditions under which self-location can arise despite the absence of an explicitly provided body. Immersive video provides a privileged case for examining this second domain.

Conceptually, self-experience in immersive video can be summarised as follows:

Self in immersive video
= dominant self-location
+ retained viewpoint-directed agency
+ constrained environment-directed agency
+ unstable ownership
+ reduced body-schema availability

This is not a quantitative predictive model, but a conceptual model showing how self-experience in immersive video is reconfigured among agency, ownership, body schema, and self-location. The model reconnects phenomenological and cognitive-scientific discussions of the minimal self, VR research on presence and the sense of embodiment, and research on multisensory integration involving visual information, vestibular sensation, and proprioception, under the media condition of non-interactive immersive video.

This model, however, also has limitations. First, the argument of the present paper is a theoretical account based on phenomenological observation and has not yet been experimentally verified. Second, the extent to which location at a bodiless viewpoint can be stably established may be strongly affected by individual differences. Some experiencers may have little difficulty with the absence of the body and may smoothly experience viewpoint-based presence. Others may find it difficult to maintain presence or self-location because of the absence of a body or of possibilities for action. Third, self-location in immersive video may vary depending on the type of chair, bodily posture, field of view, camera height, viewpoint movement, visual exploration, the degree of freedom of head rotation, and the affordances of the video content. In particular, when the video affords looking around but does not afford reaching or walking, presence and self-location may be more easily established, while the absence of environment-directed agency may become more explicit. These conditions should be systematically examined in future empirical research.

## V. CONCLUSION: Being There without a Body

This paper has reconsidered self-experience in immersive video through the relation between body schema and self-location. Self-experience in XR has conventionally been discussed in terms of bodily extension, the acquisition of a proxy body, or body ownership of an avatar. What becomes salient in immersive video viewed through a head-mounted display, however, is not the enhancement of embodiment through the provision of a body, but a configuration in which self-location becomes relatively dominant while body schema availability is reduced.

In this experience, the user does not freely act upon the recorded environment. Environment-directed agency, by which the user changes objects or events through walking, approaching, avoiding, grasping, or manipulating, is substantially constrained. Nor does the user stably own a body corresponding to the viewpoint. Nevertheless, the user often experiences the self as being located at a point within the video space. What is established here is not a self supported by body ownership or active intervention in the environment, but a self organised around viewpoint-based self-location.

This does not mean that the body is unnecessary in immersive video. The user's physiological body is, in fact, always present. Head movement, eye movement, postural adjustment, vestibular sensation, and proprioception also remain operative. Moreover, insofar as the field of view is updated in response to head rotation, viewpoint-directed agency is also retained. Self-experience in immersive video should therefore not be understood as the complete absence of agency or embodiment. Rather, it should be understood as an asymmetric configuration in which viewpoint-directed agency remains, environment-directed agency is constrained, body ownership is unstable, and body schema availability is reduced.

This is what the present paper calls a self-location-dominant state. It is not a state in which the body schema has disappeared. Nor is it an argument that presence should be reduced to self-location. Rather, it is a state in which, within the self-experience accompanying presence, self-location becomes relatively dominant over agency, ownership, and body schema. In this state, the occupation of a viewpoint becomes the principal cue for self-location, while bodily posture, bodily outline, centre of gravity, and possibilities for movement are less likely to function as explicit and stable cues.

This structure becomes particularly clear in scenes involving camera shake, impact, contact, or direct address. During ordinary viewing, presence and self-location are established relatively smoothly, and the body schema is transparent. When the stability of the viewpoint is disturbed, however, bodily awareness that had been backgrounded comes to the foreground. Crucially, the body is not recovered at that moment as a complete body image. The body is recalled, but is not reconstituted as a clear body with posture, outline, centre of gravity, and possibilities for movement. The body is not recovered; it returns incompletely. This incomplete return of the body provides a phenomenological clue that, in immersive video, self-location is established in relative dissociation from the body schema.

In this sense, immersive video is not merely a video format. It is a theoretical boundary case that prompts a reconsideration of the phenomenological assumption, developed against Cartesian dualism, that the self is grounded in the lived body, and of the cognitive-scientific vocabulary that has analysed the minimal self primarily through agency and ownership. The self in XR is not constituted only as an extension of the body. Nor is it the case that self-location cannot be established unless a body is provided. Immersive video shows that self-location can be established through a viewpoint even when no body is explicitly given. At the same time, disruption of the viewpoint paradoxically reveals the extent to which the body ordinarily supports self-experience.

The argument of the present paper is an attempt to connect research on presence, the sense of embodiment, and bodily self-consciousness. Research on presence has examined the conditions under which the experiencer feels that they are in the presented environment. Research on the sense of embodiment has clarified the relation among body ownership, agency, and self-location. Research on bodily self-consciousness has examined the integrative relation among bodily identification, self-location, and first-person perspective. Immersive video presents a condition in which these components are not enhanced as a whole, but are selectively retained, constrained, and redistributed. Self-experience in immersive video should therefore be understood not simply as bodyless presence, but as an asymmetric reconfiguration of the components of bodily self-consciousness.

Future research should empirically examine this self-location-dominant state. For example, immersive videos with stable camera positions could be compared with immersive videos involving camera impact or shake, and presence, self-location, body ownership, cybersickness, and the clarity of bodily recall could be measured. It should also be examined whether second-personal address strengthens self-location or, conversely, renders the absence of the body schema more salient. Furthermore, the extent to which location at a bodiless viewpoint can be established may vary depending on individual differences, bodily posture, the type of chair, field of view, camera height, viewpoint movement, the degree of freedom of head rotation, and the affordances of the video content. By systematically manipulating these conditions, future research could clarify more precisely the conditions under which presence and self-location are established in immersive video.

In conclusion, this paper defines self-experience in immersive video as a form of spatial self-location in which viewpoint-based self-location becomes dominant under conditions of relatively reduced body schema availability. This definition does not deny the body. Rather, it clarifies that the body supports self-experience by operating transparently, and that when this transparency breaks down, the body returns in an incomplete form. Being there without a body does not mean that the body is absent. It refers to a distinctive self-experience in immersive video in which the self is located in space through a viewpoint, even though the body is not explicitly given or stably owned. Yet if a sense of "here" can persist without stable body ownership, a more fundamental question remains: what makes this "here" an experience for someone at all—a question explored in *Present After Presence* (Toida, 2026).

## ACKNOWLEDGMENT

The author acknowledges all past and present collaborators whose discussions and intellectual exchanges have shaped the broader perspective from which this work was developed. The author would also like to thank Dr Bruno Herbelin (EPFL) for his generous and thoughtful comments on bodily self-consciousness, self-location, and presence in virtual reality during the preparation of the revised version of this manuscript. This exchange helped the author clarify the relation between immersive video, self-location, and reduced bodily contribution. Any remaining interpretations and errors are solely the author's own.